\documentclass[sigconf]{acmart}
\renewcommand\footnotetextcopyrightpermission[1]{} 
\usepackage{booktabs} 

\usepackage{tabularx}
\usepackage{subfig}
\usepackage{listings}
\usepackage{datetime}
\usepackage{flushend}

\newcommand{\URIR}{\mbox{URI-R} }
\newcommand{\URIM}{\mbox{URI-M} }

\newcommand{\URIP}{\mbox{URI-P} }

\newcommand{\URIRs}{\mbox{URI-Rs} }
\newcommand{\URIMs}{\mbox{URI-Ms} }

\newcommand{\URIMns}{\mbox{URI-M}}
\newcommand{\URIRns}{\mbox{URI-R}}

\newcommand{\URIPns}{\mbox{URI-P}}

\newcommand{\URIRsns}{\mbox{URI-Rs}}
\newcommand{\URIMsns}{\mbox{URI-Ms}}
\newcommand{\URITsns}{\mbox{URI-Ts}}
\newcommand{\URIGsns}{\mbox{URI-Gs}}

\usepackage{array}
\usepackage{bold-extra}
\newcolumntype{L}{>{$}l<{$}} 
\newcolumntype{R}{>{$}r<{$}} 
\newcommand{\arr}{$ \rightarrow $}
\usepackage{makecell}

\newcommand{\entity}{mementity }
\newcommand{\entityns}{mementity}

\newcommand{\entities}{mementities }
\newcommand{\entitiesns}{mementities}

\newtoggle{showfeedback}
\toggletrue{showfeedback}

\newcommand{\feedbackComment}[1]{}
\newcommand{\feedbackFootnote}[1]{}
\newcommand{\footnoteFeedback}[1]{}
\newcommand{\feedbackMargin}[1]{}

\makeatletter
\newcommand\setItemnumber[1]{\setcounter{enum\romannumeral\@enumdepth}{\numexpr#1-1\relax}}
\makeatother

\usepackage{tikz}
\newcommand{\blackcircle}[2][black,fill=black]{\tikz[baseline=-0.5ex]\draw[#1,radius=#2] (0,0) circle ;}%
\usepackage{mdframed}

\usepackage{enumitem}
\usepackage{multirow}

\newcommand{\specialcell}[2][c]{%
  \begin{tabular}[#1]{@{}c@{}}#2\end{tabular}}

\begin{document}

\fancyhead{}
\title{A Framework for Aggregating Private and Public Web Archives}

\author{Mat Kelly}
\affiliation{%
  \institution{Old Dominion University}
  \city{Norfolk}
  \state{Virginia} 
  \postcode{23529}
  \country{USA}
}
\email{mkelly@cs.odu.edu}

\author{Michael L. Nelson}
\affiliation{%
  \institution{Old Dominion University}
  \city{Norfolk} 
  \state{Virginia} 
  \postcode{23529}
  \country{USA}
}
\email{mln@cs.odu.edu}

\author{Michele C. Weigle}
\affiliation{%
  \institution{Old Dominion University}
  \city{Norfolk} 
  \state{Virginia} 
  \postcode{23529}
  \country{USA}
}
\email{mweigle@cs.odu.edu}

\begin{abstract}
Personal and private Web archives are proliferating due to the increase in the tools to create them and the realization that Internet Archive and other public Web archives are unable to capture personalized (e.g., Facebook) and private (e.g., banking) Web pages. We introduce\footnote{This is a preprint version of the ACM/IEEE Joint Conference on Digital Libraries (JCDL 2018) full paper: \url{https://doi.org/10.1145/3197026.3197045}.} a framework to mitigate issues of aggregation in private, personal, and public Web archives without compromising potential sensitive information contained in private captures. We amend Memento syntax and semantics to allow TimeMap enrichment to account for additional attributes to be expressed inclusive of the requirements for dereferencing private Web archive captures. We provide a method to involve the user further in the negotiation of archival captures in dimensions beyond time. We introduce a model for archival querying precedence and short-circuiting, as needed when aggregating private and personal Web archive captures with those from public Web archives through Memento. Negotiation of this sort is novel to Web archiving and allows for the more seamless aggregation of various types of Web archives to convey a more accurate picture of the past Web.
\end{abstract}


\maketitle

\section{Introduction}

Conventional Web archives preserve publicly available content on the live Web. Some Web archives allow users to submit URIs to be individually preserved or used as seeds for an archival crawl. However, some content on the live Web may be inaccessible (e.g., beyond the crawler's capability compared to a live Web browser) or inappropriate (e.g., requires a specific user's credentials) for these crawlers and systems to preserve. For this reason and enabled by the recent influx of personal Web archiving tools, such as WARCreate, WAIL, and Webrecorder.io, individuals are preserving live Web content and personal Web archives are proliferating \cite{marshall-rethinking}. 

Personal and private captures, or mementos, of the Web, particularly those preserving content that requires authentication on the live Web, have potential privacy ramifications if shared or made publicly replayable after being preserved \cite{Marshall:2012:IAS:2232817.2232819}. 
Given the privacy issues, strategically regulating access to these personal and private mementos would allow individuals to preserve, replay, and collaborate in personal Web archiving endeavors. Adding personal Web archives with privacy considerations to the aggregate view of the ``Web as it was'' will provide a more comprehensive picture of the Web while mitigating privacy violations.

This work has four primary contributions to Web archiving:
\begin{flushleft}
\textbf{Archival Query Precedence and Short-circuiting}: Allow querying of individual or subsets of archives of an aggregated set in a defined order with the series halting if a condition is met (Section~\ref{sec:precedence}).\\
\textbf{TimeMap/Link Enrichment}: Provide additional, more descriptive attributes to \URIMs for more efficient querying and interaction (Section~\ref{sec:additionalTimeMapAttributes}).\\
\textbf{Multi-dimensional user-driven content negotiation of archives}: Increase user involvement in request for \URIMs in both temporal and other dimensions (Sections~\ref{sec:mma} and \ref{sec:negotitationDimensions}).\\
\textbf{Public/Private Web Archive Aggregation}: Introduce additional special handling of access to private Web archives for Memento aggregation using OAuth (Section~\ref{sec:auth}).
\end{flushleft}

%


\begin{figure}
\captionsetup[subfigure]{justification=centering}
\centering
\subfloat[Local Archive capture of \url{facebook.com}][Local Archive capture\\ of \url{facebook.com}]{
\includegraphics[width=0.45\linewidth]{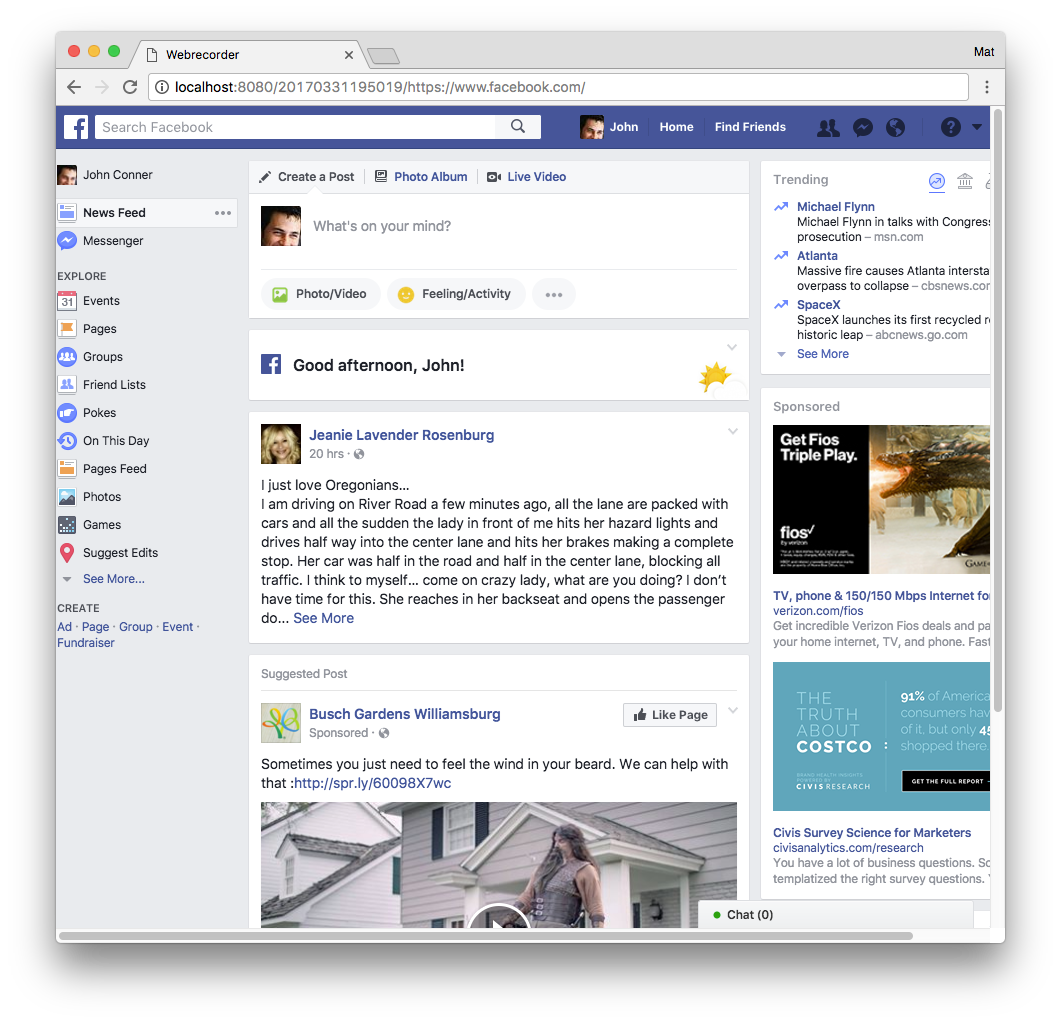}
\label{fig:fb_me}
}\hspace{1.0em}
\subfloat[Internet Archive capture of \url{facebook.com}][Internet Archive capture\\of \url{facebook.com}]{
\includegraphics[width=0.45\linewidth]{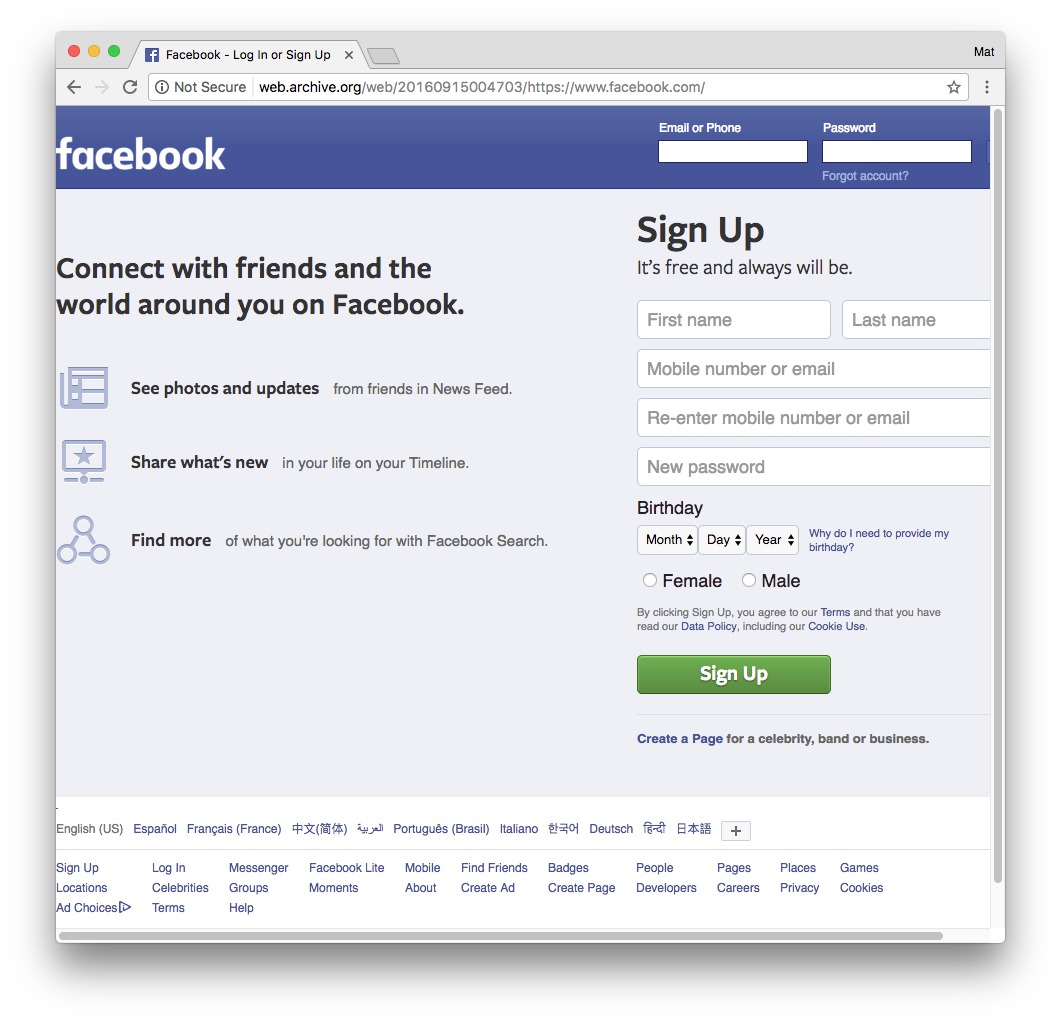}
\label{fig:fb_ia}
}\\
\subfloat[Private content on the live Web that is extremely time sensitive to preserve for future access.]{
\includegraphics[width=1.0\linewidth]{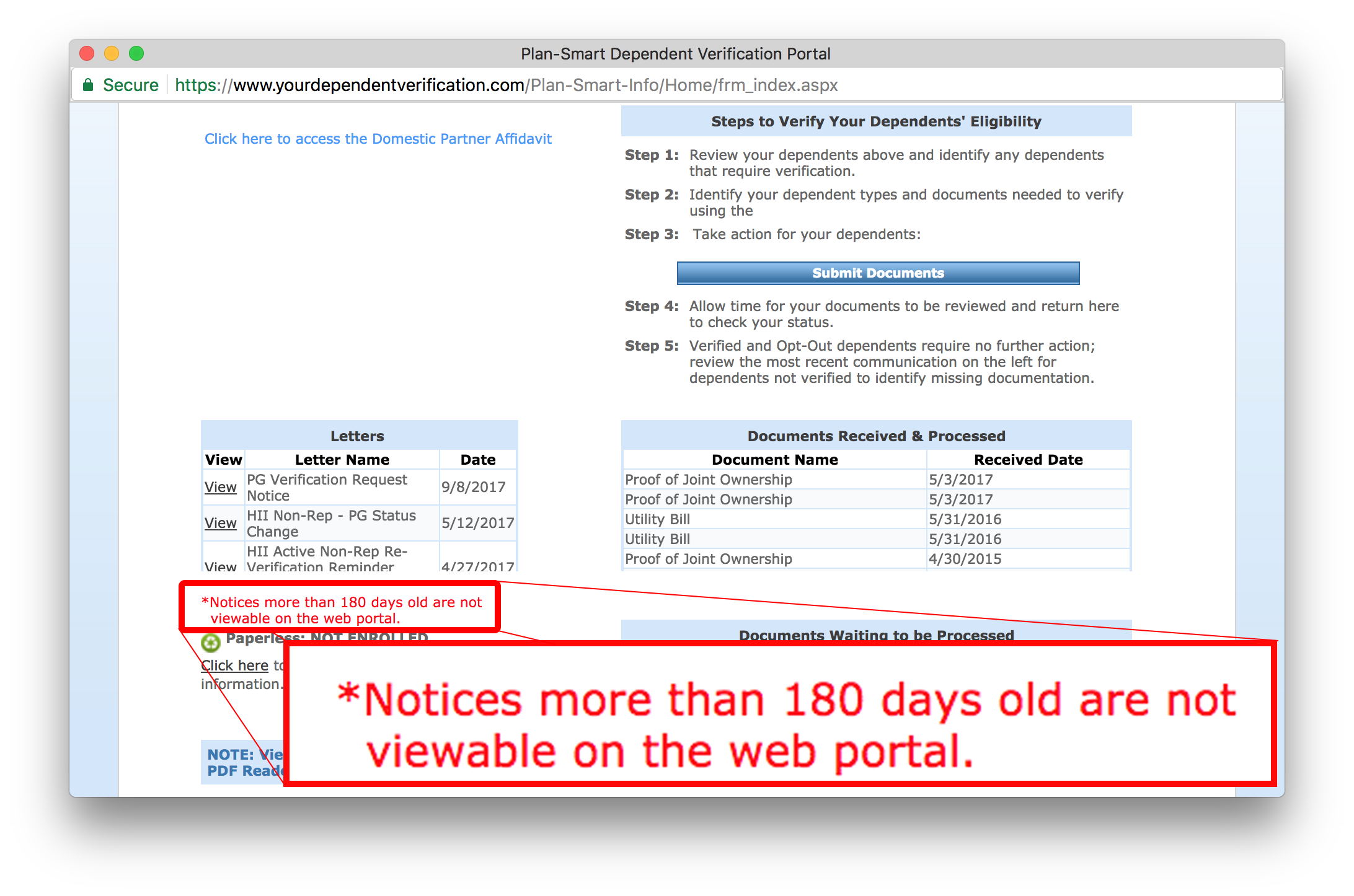}
\label{fig:hii}
}
\caption[test]{Personalized and Private Web pages.}
\label{fig:fb}
\end{figure}

\subsection{Solutions Beyond Institutions}

Personal Web archives may contain captures with personally identifiable information, such as a time sensitive statement verification Web page (Figure~\ref{fig:hii}) or a user's \texttt{facebook.com} feed (Figure~\ref{fig:fb_me}). A user may want to selectively share their \url{facebook.com} mementos \cite{1555440} but wish to also regulate access to them \cite{Marshall:2014:AAF:2740769.2740772}. Without the ability of authenticating as a user on the live Web, many public Web archives simply preserve the \url{facebook.com} login page (Figure~\ref{fig:fb_ia}). Both captures are representative of \texttt{facebook.com}, and they may have even been captured at the same time. Users may be hesitant to share their mementos of \url{facebook.com} (or other personal or private Web pages) without a mechanism to ensure that the Web page as the user experienced it is faithfully captured and that the access of those captures can be regulated.

As a counterpoint, an individual's personal Web archive is more susceptible to disappearing without an institution's backing. Maintaining backups of archived content is unwieldy, requires diligence or automation, and is still at the mercy of hardware failures. While distributed propagation of the captures to other places may ameliorate this issue, another privacy issue remains in that distributed content may be sensitive and must be handled differently at the level of access. 

\begin{figure}
\centering
\includegraphics[width=1.0\linewidth]{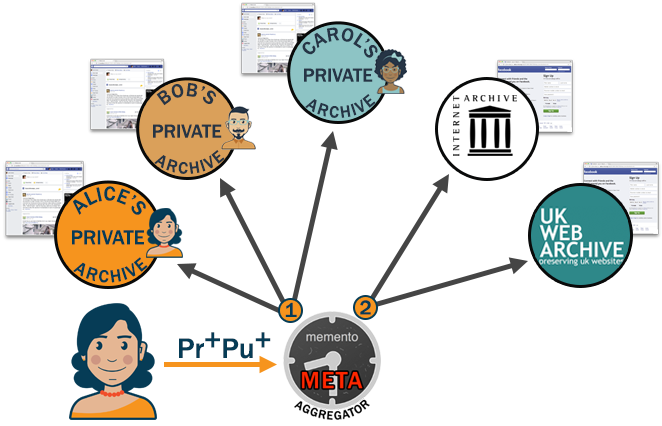}
\caption{Archival precedence using private first then public Web archiving querying model (Pr$^+$Pu$^+$).}
\label{fig:precedence}
\end{figure}

\begin{figure}
\centering
\includegraphics[width=1.0\linewidth]{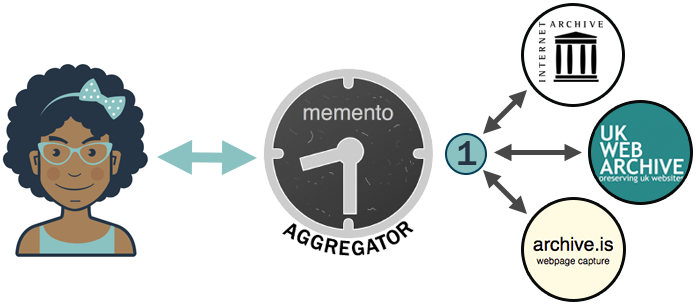}
\caption{Conventional Memento aggregators query a set of public Web archives and do so in an equally-weighted querying model \-circuiting.}
\label{fig:ma}
\end{figure}

To observe the more representative picture of ``what I saw on the Web of the past'' inclusive of private Web archive captures, we could give precedence to private Web archives over public Web archives when aggregating. For example, temporally aggregating my friends' captures (potentially residing in multiple private Web archives) with those consisting of preserved Facebook login pages (Figure~\ref{fig:fb_me}) from public Web archives (who are rightly not responsible for preserving my Facebook feed) may not be desirable. Instead, a user may want to instruct the aggregator to only aggregate mementos from archives with certain characteristics, e.g., a set of private Web archives, and only if no personal captures are found, look to the public Web archives for captures (Figure~\ref{fig:precedence}). This sort of runtime specification of archival query precedence does not currently exist. Today's Memento aggregators can only query a static set of public Web archives specified at configuration time (Figure~\ref{fig:ma}).

Because more personal and private, non-institutional backed Web archives are being created, and these archives may contain sensitive data that cannot be shared without special handling, more work must to be done to address the impermanence of personal Web archives with consideration for their contents.


\subsection{Enrichment of Archival Aggregates}

We provide amendments to the semantics of Memento TimeMaps to encourage aggregation of mementos from more archives while still allowing for the distinction between conventional and enriched captures with additional metadata. We introduce additional \textit{\entities}(a portmanteau of ``Memento'' and ``entity'')\footnote{Used for distinction from the term ``entity'' as defined and in the now-deprecated RFC2616 describing HTTP/1.1.} for accessing various types of Web archives. The use of \entities could enable negotiation in additional dimensions beyond time, systematic aggregation of private captures, regulated access control to Web archives that may contain personal or private mementos, etc.

\begin{figure*}[t]
\begin{lstlisting}
!context ["https://oduwsdl.github.io/contexts/memento"]
!id {"uri": "http://localhost:1208/timemap/cdxj/http://facebook.com"}
!keys ["memento_datetime_YYYYMMDDhhmmss"]
!meta {"original_uri": "http://facebook.com"}
!meta {"timegate_uri": "http://localhost:1208/timegate/http://facebook.com"}
!meta {"timemap_uri": {"link_format": "http://localhost:1208/timemap/link/http://facebook.com", "json_format": "http://localhost:1208/timemap/json/http://facebook.com", "cdxj_format": "http://localhost:1208/timemap/cdxj/http://facebook.com"}}
19981212013921 {"uri": "http://archive.is/19981212013921/http://facebook.com/", "rel": "first memento", "datetime": "Sat, 12 Dec 1998 01:39:21 GMT"}
19981212013921 {"uri": "http://web.archive.org/web/19981212013921/http://facebook.com/", "rel": "memento", "datetime": "Sat, 12 Dec 1998 01:39:21 GMT"}
19981212024839 {"uri": "http://web.archive.org/web/19981212024839/http://www.facebook.com/", "rel": "memento", "datetime": "Sat, 12 Dec 1998 02:48:39 GMT"}
...
20170330231113 {"uri": "http://web.archive.org/web/20170330231113/http://www.facebook.com/", "rel": "memento", "datetime": "Thu, 30 Mar 2017 23:11:13 GMT"}
20170331013527 {"uri": "http://web.archive.org/web/20170331013527/https://www.facebook.com/", "rel": "last memento", "datetime": "Fri, 31 Mar 2017 01:35:27 GMT"}
\end{lstlisting}
\caption{An abbreviated CDXJ TimeMap from MemGator for \url{facebook.com}.}
\label{fig:cdxj}
\end{figure*}

In this work (Section~\ref{sec:additionalTimeMapAttributes}) we introduce three new types of attributes for richer TimeMaps: \textbf{content-based attributes} based on data when a URI is dereferenced, \textbf{derived attributes} requiring further analysis beyond dereferencing but useful for evaluating capture quality, and \textbf{access attributes} that guide users and software as to requirements needed to dereference mementos in private, personal, and archives with access restrictions.

Through this TimeMap\footnoteFeedback{MLN: and Link header!} enrichment, a user will be able to specify the semantics to be selective in the set of archived URIs (\URIMsns) returned in a TimeMap through a set of attributes beyond time and original URI (\URIRns). This will allow the user to interact with the Memento aggregator to specify a custom subset and/or supplement the existing supporting archives in the aggregated result returned.

Conventional Memento aggregation \cite{sanderson-global} (Figure~\ref{fig:ma}) is accomplished by a user requesting captures for a \URIR from a remote endpoint. The software receiving the request then relays this request to a set of Web archives with which it is configured. Once the Web archives return a response containing their captures, the aggregator software temporally sorts the \URIMsns, adds additional Memento metadata, and returns the aggregated TimeMap to the user. 

In this work we also describe a cascading hierarchical relationship between the aggregators and the \entities involved in aggregation and negotiation. We introduce a ``Meta-Aggregation'' concept (Section~\ref{sec:mma}) to allow for a recursive relationship of one aggregator onto another, ``building up'' an aggregate result, potentially including supplemental information in the aggregation. Section~\ref{sec:archivalSelection_mma} describes multiple scenarios where supplementing the memento aggregation using a meta-aggregator \entity would be useful. This hierarchy may also include other \entities like one to regulate access to private Web archives (identified with a \URIPns) and another, a StarGate, to allow for selective negotiation (Section~\ref{sec:stargate}), e.g., allowing the client to request that only results from private Web archives are returned from an aggregator. 

\section{Background and Related Work}
\label{sec:bg}
\label{sec:relatedWork}

\subsection{Archiving and Linked Data}

The Memento Framework \cite{rfc7089} provides the constructs to interact with Web archives in the temporal dimension. An archival capture (memento) is identified by a \URIMns. Memento aggregation allows identifiers for mementos (\URIMsns) from multiple Web archives to be temporally sorted using the parameters of the original live Web URI (\URIRns) and a timestamp of the capture (Memento-Datetime). Memento TimeMaps contain a listing of \URIMs for mementos of an original resource on the live Web. TimeMaps also include contextual information like the \URIR that the TimeMap represents, URIs for a Memento \entity to handle temporal content negotiation (TimeGate, i.e., \URIGsns), and identifiers for other TimeMaps (\URITsns). 

Memento TimeMaps are conventionally formatted and extend upon the Web Linking specification \cite{rfc8288}. The syntax of the Link format applies to both information expressed in HTTP headers as well as information supplied in a TimeMap listing. Because of this, a limited set of attributes about \URIMs is allowed within a TimeMap inclusive of \texttt{rel} and \texttt{datetime}. Additional information about a \URIM would be useful if present in a TimeMap. For example, knowing the HTTP status code of the dereferenced \URIM would reduce the amount of time needed to determine unique captures in the archive \cite{kelly-jcdl2017}. 
Extending TimeMaps may also provide the facility for the integration of private and public Web archives. 

Alam et al. \cite{salam-cdxj} defined the CDXJ format, an extension of the conventional CDX \cite{cdx} archival indexing format, as an extensible means of associating additional attributes to \URIMsns. CDX files serve as indexes for Web archive files and contain many fields, like MIME-type, status code, and content-digest of the memento, which are not present in TimeMaps. MemGator \cite{memgator} is an open source Memento aggregator that supports CDXJ TimeMaps (example in Figure~\ref{fig:cdxj}) along with conventional Link-formatted and additionally JSON-formatted TimeMaps. In this work, we adapt the code for MemGator to handle additional HTTP request parameters supplied by a client as well as producing TimeMaps with the additional proposed attributes. 

While these two dimensions are sufficient for the aggregation of public Web archives, additional parameters are required to express the need for privacy considerations or further steps to be executed to dereference the \URIMns. Beyond the ability to express distinction in private and public mementos, it may not make sense to request public Web archive captures from an aggregator based on a variety of conditions. Some examples where expressions to distinguish captures are in isolation by URI (explicit exclusion of public captures from results) and Archival precedence (e.g., only check for captures of facebook.com in public Web archives when none are in my own).

The `profile' link relation type \cite{rfc6906} (discussed in Section~\ref{sec:precedence}) provides a standard set of semantics for processing a resource representation. We leverage and extend these semantics to allow a user to request mementos with certain properties from Web archives.

In earlier work, we developed WARCreate \cite{kelly-jcdl12}, a Google Chrome browser extension, to allow a user to capture content from their browser, even pages behind authentication, into the standard web archiving (WARC) format. We also re-packaged institutional grade Web archiving tools in WAIL \cite{berlin-wail}, a native desktop application, to allow individuals to preserve, replay, and retain complete control of their captures. More recently, the Webrecorder.io service allows similar capability, including allowing the user to capture content behind authentication. But unlike WARCreate and WAIL, Webrecorder relies on the user's credentials being proxied through the service, a potentially undesirable feature with privacy ramifications. 

\subsection{Privacy and Security}

The Snowden Archive-in-a-Box project \cite{snowden} is an autonomous version of the the Snowden Digital Surveillance Archive. The project uses a Raspberry Pi single-board computer along with other hardware and a data set containing files leaked by Edward Snowden to allow browsing of the files without a user being surveilled. This use case highlights access as being the problematic factor beyond the base case of the content being sensitive. When aggregating captures from a Snowden archive with captures from other archives, requesters may wish to prevent requests from propagating to other archives via the aggregator (by specifying a privateOnly profile) or to only consult other archives when no results are returned from their instance (request precedence, both discussed in Section~\ref{sec:precedence}).

While little research has been performed on the aggregation of private and public captures, multiple surveys have been performed by a variety of researchers on user's perspectives on private Web archives. In particular, Marshall and Shipman \cite{Marshall:2012:IAS:2232817.2232819} surveyed Web users on potential efforts for institutions to preserve their private Web contents. They particularly highlighted the need for exploration of who retains control of access of private content once it is preserved and made available.

The OAuth 2.0 Authorization Framework \cite{rfc6749} (usage discussed in Section~\ref{sec:accessAttributes}) provides a model for tokenization that we apply for persistent access to private Web archives. Using this model requires a secondary authorization server, implemented in this research as an additional \entity to decouple the authentication burden on the archive. This model, however, requires the archive to be aware of this additional \entity to act as a gateway to the archive.

Cushman and Kreymer \cite{ilya-hacker} performed an extensive review on the security of Web archives in the context of both preservation and replay. Through technical examples on how an attacker might capture private resources, they provided approaches for mitigating each sort of attack. 
In related work, Brunelle et al. \cite{brunelle-dlib2016} described issues with private Web archiving on an organizational scale. Through analyzing the results of an archival crawler instance, they identified content that should not have been accessible to the crawler, which required wholesale removal of WARCs containing the information for lack of a method of selective removal. 

\subsection{Memento and HTTP Mechanics}

Rosenthal \cite{dshr-aggregators} emphasized that temporal order may not be optimal for TimeMaps returned from Memento aggregators. He stated that aggregators need to develop ways of estimating usefulness of preserved content and conveying these estimates to readers. In a different work, Rosenthal \cite{dshr-importance} described the behavior of aggregators returning ``Soft 403s'' consisting of captures of login pages when the user likely expected content shown that was originally behind authentication. 

Rosenthal \cite{dshr-importance} also described a ``hints list'' that an aggregator might provide based on its own experience of requesting content from archives. In this work Rosenthal also alluded to a hypothetical mechanism of the aggregator filtering content like login pages from the results and redirecting a user to a version of the TimeMap containing only captures that are not a login page.

Jones et al. \cite{sjones-raw1, preferHeader-wsdlBlog} discussed obtaining the ``raw mementos'' consisting of un-rewritten links in captures in a systematic way using the HTTP Link response header. By utilizing the HTTP Prefer request header \cite{rfc7240}, a user would be able to obtain a version of the memento as it appeared at the time of capture instead of a version with relative links rewritten by the archive to point back within the archive and not the live Web. An archive, in response and to confirm compliance with the request, would return the memento with the HTTP Preference-Applied response header along with the requested original version of the memento.

The HTTP Prefer header \cite{rfc7240} allows an explicit means for a client to express their preferences of optional aspects in an HTTP Request. Van de Sompel et al. \cite{preferHeader-wsdlBlog} highlighted that the Prefer header could be used by Web archives to allow clients to specify a request for the unaltered or un-rewritten content. Rosenthal \cite{preferHeader-dshr} echoed Van de Sompel et al. by suggesting a list of transformations (screenshot, altered-dom, etc.) for a memento via a new HTTP header. 

This work focuses on the transformation of TimeMaps, not the mementos themselves. The rewriting problem in previous work is pertinent to replay of \URIMs whereas what we accomplish is more expressive metadata of the mementos prior to and to mitigate issues with dereferencing \URIMsns. A goal of this work is to further involve the client in the aggregation process. Interaction with the aggregators through these sort of mechanisms will be a first step in accomplishing this.

Fielding and Reschke \cite{rfc7231} defined proactive and reactive content negotiation as that which is determined by the server as a best-guess based on metadata and a model involving communication and selection of representations, respectively. The latter may be accomplished a variety of ways inclusive of the utilization of the HTTP \texttt{300 Multiple Choices} and \texttt{406 Not Acceptable} status codes as well as the less commonly implemented HTTP Alternates header and Transparent Content Negotiation \cite{rfc2295}. As we anticipate generating derivative TimeMaps consisting of any number of permutations of additional attributes applied on the mementos, it would be useful to associate and allow users to choose the variant they prefer using these status codes and HTTP transaction patterns.

In previous work \cite{kelly-msthesis}, we highlighted an issue of URI-collision in the realm of personal Web archives wherein (for example) both a login page and the authenticated content of a live Web application may reside at the same \URIR (Figure~\ref{fig:fb}). We \cite{kelly-dlib2013} extended this work by identifying personalized representations of mementos and providing a mechanism to navigate between additional dimensions beyond time. As personal Web archives proliferate and are at some point aggregated into multi-archive TimeMaps (cf. a TimeMap from and containing only listings from the archive itself), it would be useful to distinguish \URIMs that represent personalized mementos, mementos that were originally behind authentication, and mementos in personal Web archives that require additional considerations and mechanisms to access.

\section{Archive Query Precedence and Short-Circuiting}
\label{sec:precedence}

Private Web archives contain an inherent characteristic where exposing the metadata about an archive's contents could be sufficient to identify the archive's contents. For example, a private archive responding with a TimeMap containing \URIMs for captures of my online bank statement would reveal where I am banking as well as where I am preserving personal banking information.


To mitigate the unnecessary revelation of potentially personal information, a client who has set up a Memento aggregator with access to their private Web archive may wish to have requests sent to public Web archives only if no results are returned from their private Web archive.

Figure~\ref{fig:precedence} illustrates requests being first sent to the private archives then to public Web archives. But, it may also be desirable to allow this type of behavior to functionally coexist with conventional pipelined asynchronous archive querying. As with the Snowden Archive-in-a-Box example in Section~\ref{sec:bg}, checking for the existence of captures of this content in other archives may imply interest or association with the subject matter, in some cases itself being revealing or even incriminating. To maintain privacy, a Memento aggregator with access to the Snowden archive would require special handling of requests for resources that might be contained in that archive. For example, a user may want requests for a certain \URIR to only be requested from the Snowden archive or their own personal Web archive, and not other public or private Web archives. 

We propose two initial approaches to accomplish this: explicit specification by a client at the time of request and analysis of mementos with a potentially personalized representation. For the latter, we \cite{kelly-dlib2013} identified three methods for identifying personalized representations. Of the methods proposed, but not investigated further (we opted for one of the other three), was to be able to specify additional environment variables when selecting a representation of a resource. The downside, we mentioned, was the requirement of a specialized client. The specialized ``client'' in this case may be the \entity responsible for determining the degree of personalization of the representation, i.e., the StarGate.

When aggregating and replaying a \URIR over time from a set of archives consisting of captures from both public and private Web archives, it may be desirable to first check for private captures prior to requesting \URIMs from public Web archives (Figure~\ref{fig:precedence}). For example, in aggregating \URIMs for \texttt{facebook.com} that include mementos of my news feed from my private archive and unauthenticated login pages from institutional public Web archives  (Figure~\ref{fig:fb}), the latter is less useful in observing how the page has changed over time. To maintain relevancy of the desired sort of representation, we may want to check for the existence of captures from private Web archives \textit{first} and then, only if none are present, resort to requesting the captures consisting of a login page. This model of precedence (request priority) and short-circuiting (stop requesting captures if a condition is met) via Memento aggregators does not currently exist but could be critical in a user expressing what they expect from an aggregator beyond simply mementos for a \URIRns. 

\begin{figure}
\centering
\includegraphics[width=1.0\linewidth]{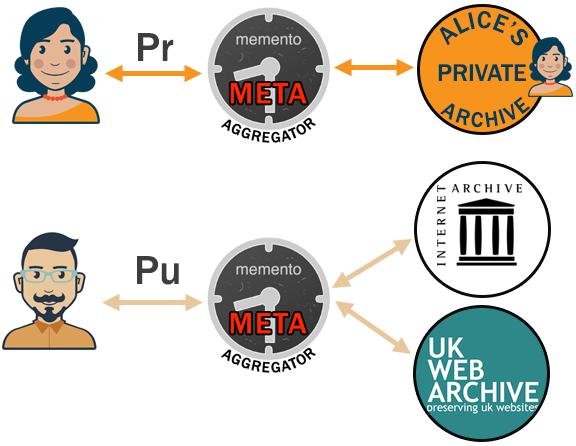}
\caption{PrivateOnly and PublicOnly aggregation in an MMA.}
\label{fig:mma_prpu}
\end{figure}

In the basic model below, we express various access precedence models (henceforth \textit{profile}) for containing boolean categorization of private and public Web archives. In each profile, order is significant and thus a simple regular expression can be used where $P_u$ symbolizes a public Web archive endpoint, $P_r$ a private Web archive endpoint, and the ``+'' superscript indicating at least one or more consecutive instances.
\vspace{2.3em} 
\begin{equation}\label{eqn:noArchives}noArchives \rightarrow \varnothing \rightarrow \{\}\end{equation}
\begin{equation}\label{eqn:publicOnly}publicOnly \rightarrow {P_u}^+\end{equation}
\begin{equation}\label{eqn:privateOnly}privateOnly \rightarrow {P_r}^+\end{equation}
\begin{equation}\label{eqn:privateFirst}privateFirst \rightarrow {P_r}^+{P_u}^+\end{equation}
\begin{equation}\label{eqn:publicFirst}publicFirst \rightarrow {P_u}^+{P_r}^+\end{equation}

The basic profiles pair with the syntax of the \texttt{profile} relation type \cite{rfc6906}, allowing clients to request resulting TimeMaps containing \URIMs from a subset of archives from which the Memento \entity requests (Figure~\ref{fig:mma_prpu}). The preliminary scheme for short-circuiting of subsequent requests is also boolean, e.g., requests should only be made to public Web archives when the \texttt{privateFirst} profile (Equation~\ref{eqn:privateFirst}) is specified by the client when no identifiers for captures are returned from private archives. This model also assumes that the sets $P_u$ and $P_r$ are disjoint ($P_u \cap P_r = \varnothing$) for simplicity, but may not be the case in reality. For Web archives that contain both private and public captures, an approach toward achieving mutually exclusivity could be to separate each set of the private and public \URIRs into an abstraction of separate collections. For example, as discussed earlier, the UK Web Archive contains captures from its legal deposit with restricted off-site access; that is, a user cannot access the mementos unless physically on location at the library (Figure~\ref{fig:bl_451_curl}).

\begin{figure}
\begin{mdframed}[leftmargin=1em,rightmargin=1em]
\begin{flushleft}{\ttfamily\scriptsize\color{black}
\$ curl -I https://www.webarchive.org.uk/wayback/archive/*/http://www.example.org\\
\textbf{\color{red}HTTP/1.1 451 Unavailable For Legal Reasons}\\
Date: Wed, 25 Oct 2017 04:39:35 GMT\\
Server: Apache-Coyote/1.1\\
Content-Type: text/html;charset=utf-8\\
Transfer-Encoding: chunked\\
Set-Cookie: JSESSIONID=823BD09DF8DD489087763640A8150023; Path=\/; HttpOnly\\
Content-Language: en\\
}
\end{flushleft}
\end{mdframed}
\caption{Accessing a \URIM at UKWA using curl returns an HTTP 451 status code.}
\label{fig:bl_451_curl}
\end{figure}

\section{Additional TimeMap Attributes}
\label{sec:additionalTimeMapAttributes}

Aggregating private and public mementos requires the ability to distinguish captures that require special handling. To accomplish this and to provide the ability for TimeMaps to be more descriptive of the \URIMs they contain, we extend the TimeMap syntax and semantics to allow additional attributes. In this work we introduce three new types of attributes for richer TimeMaps: \textbf{content-based attributes} based on data when dereferenced, \textbf{derived attributes} requiring further analysis beyond dereferencing but useful for evaluating capture quality, and \textbf{access attributes} that guide users and software as to requirements needed to dereference mementos in private, personal, and archives with access restrictions. In this work we focus primarily on the access attributes but define other classes of attributes for future extensibility. 

\subsection{Content-based Attributes}
\label{sec:contentBasedAttributes}
In previous work \cite{kelly-jcdl2017}, we highlighted that the \URIMs in a TimeMap for \texttt{google.com} produce nearly 85\% HTTP redirects when dereferenced. Determining how many mementos exist from an archive for a \URIR is thereby impossible from a TimeMap alone. Enriching a TimeMap with information about the dereferenced capture would improve methods for determining how well (in quantity) a \URIR has been captured. HTTP data obtained when dereferencing a \URIM like status code \cite{rfc7231}, content-type \cite{rfc7231}, and Last-Modified \cite{ainsworth-hypertext2015} would be useful. For some of these attributes (like the aforementioned), the data may exist in the archival indexes, typically formatted as CDX. However, while many Web archives expose a Memento endpoint, few make these additional content-based attributes about the captures available through a CDX server\footnote{Internet Archive does currently expose a CDX endpoint with limited fields at \protect\url{http://web.archive.org/cdx/search/cdx?url=example.com}}. Thus, once these attributes are discovered by dereferencing, they may be retained and expressed with the assumption that they are an accurate account of the archival record.

\subsection{Derived Attributes}
\label{sec:derivedAttributes}
Other attributes about a memento may require calculation to obtain, which can be computationally and temporally expensive when performed at archival or even \URIR scale. This section briefly describes one such example of a derived attribute: Memento Damage calculation. Adding the ability for this derived attribute to be present in a TimeMap would allow for more efficient evaluation of memento quality. Brunelle et al. \cite{brunelle-ijdl2015-damage} developed a metric for determining the quality of a capture (cf. quantity in Section~\ref{sec:contentBasedAttributes}) when dereferencing a \URIM with a particular focus on the quantitative significance of missing embedded resources. Determining ``Memento Damage'' requires calculation beyond simple counting, as all resources are not equally weighted, particularly when absent. Having this information calculated and present in a TimeMap would allow a user to select the best or most complete \URIM without needing to iterate through all \URIMsns.

\subsection{Access Attributes}
\label{sec:accessAttributes}
An impetus for this research is integrating private and personal Web archives through aggregation via Memento TimeMaps. CDXJ allows additional attributes to be specified and considered when a \URIM is dereferenced. Figure~\ref{fig:cdxj_new} shows how a token may be stored in an enriched CDXJ TimeMap where the authentication procedure is discoverable at runtime. We utilize OAuth2 \cite{rfc6749} for authorization when dereferencing URI-Ms with this field for tokenization for persistent access to private mementos.

Access control may be needed in cases where private and personal Web archives are aggregated with public Web archives via TimeMaps. An authentication procedure and subsequent tokenization will allow persistent access using a token derived from authenticating. Figure~\ref{fig:cdxj_new} shows a token being attributed on a per-\URIM basis, though a single token may be applied to all \URIMs returned from an archive. The responsibility for attributing the token to an individual or set of mementos may lie in either the archive itself or the aggregator in this preliminary model.


\subsection{Sources of Derivatives}
CDXJ allows metadata fields (lines beginning with \texttt{!meta}) about the TimeMap to precede the listing of captures. Figure~\ref{fig:cdxj} contains metadata fields within a CDXJ TimeMap that are typically also found in a Link-formatted TimeMap, e.g., \URIR for the original resource, TimeGates, and other related TimeMaps. With the introduction of derived attributes (Section~\ref{sec:derivedAttributes}), it is critical to not just give context as to the semantics of new attributes like ``damage'' but also to provide guidance in generating this value. 

Figure~\ref{fig:cdxj_new} provides an example where a derived attribute requiring calculation (memento damage \cite{brunelle-ijdl2015-damage}) and an access attribute are defined for guidance within the TimeMap. Definitions in the \texttt{context} utilize a URI to associate semantics of an attribute for a memento. We are still currently exploring further syntax for more expressive attributes in CDXJ TimeMaps.


\begin{figure}
\begin{lstlisting}
!context ["https://oduwsdl.github.io/contexts/memento", "https://oduwsdl.github.io/contexts/damage", "https://oduwsdl.github.io/contexts/access"]
!id {
  "uri": "http://localhost:1208/timemap/cdxj/http://facebook.com"}
!meta {"...": "..."}
19981212013921 {
  "uri": "http://localhost:8080/20101116060516/http://facebook.com/",
  "rel": "memento",
  "datetime": "Tue, 16 Nov 2010 06:05:16 GMT",
  "status_code": 200,
  "damage": 0.24,
  "access": {
    "type": "Blake2b",
    "token": "c6ed419e74907d220c6647ef0a3a88a41..."
  }
}
\end{lstlisting}
\caption{An amended CDXJ record for a private capture of \texttt{facebook.com}. Line breaks added for readability.}
\label{fig:cdxj_new}
\end{figure}

\section{Memento Meta-Aggregator}
\label{sec:mma}

In this work we extend on the role of a Memento aggregator to possess additional capabilities when interacting with Web archive users, Web archives, other Memento aggregators, and other \entitiesns. MemGator \cite{memgator} allows a user to host a Memento aggregator at a location of their choosing (inclusive of the user's local machine) and configure a set of Web archives to query when starting the software. While conventional, remotely located aggregators assume that all Web archives queried are publicly accessible, a locally hosted, customizable aggregator may interface with archives that have restricted access. For example, a MemGator instance may request mementos from a Web archive that is only accessible on the user's local area network or co-hosted on the user's machine on which the aggregator resides. Non-public archives are treated and aggregated agnostically without further consideration of their holdings. In this section we describe a \entity to account for the shortcomings of conventional Memento aggregators while also extending their standard functionality and Memento interfaces.

\begin{figure*}\footnotesize
\begin{mdframed}
\centering
\begin{tabular}{ c | c }
  \includegraphics[width=0.55\linewidth]{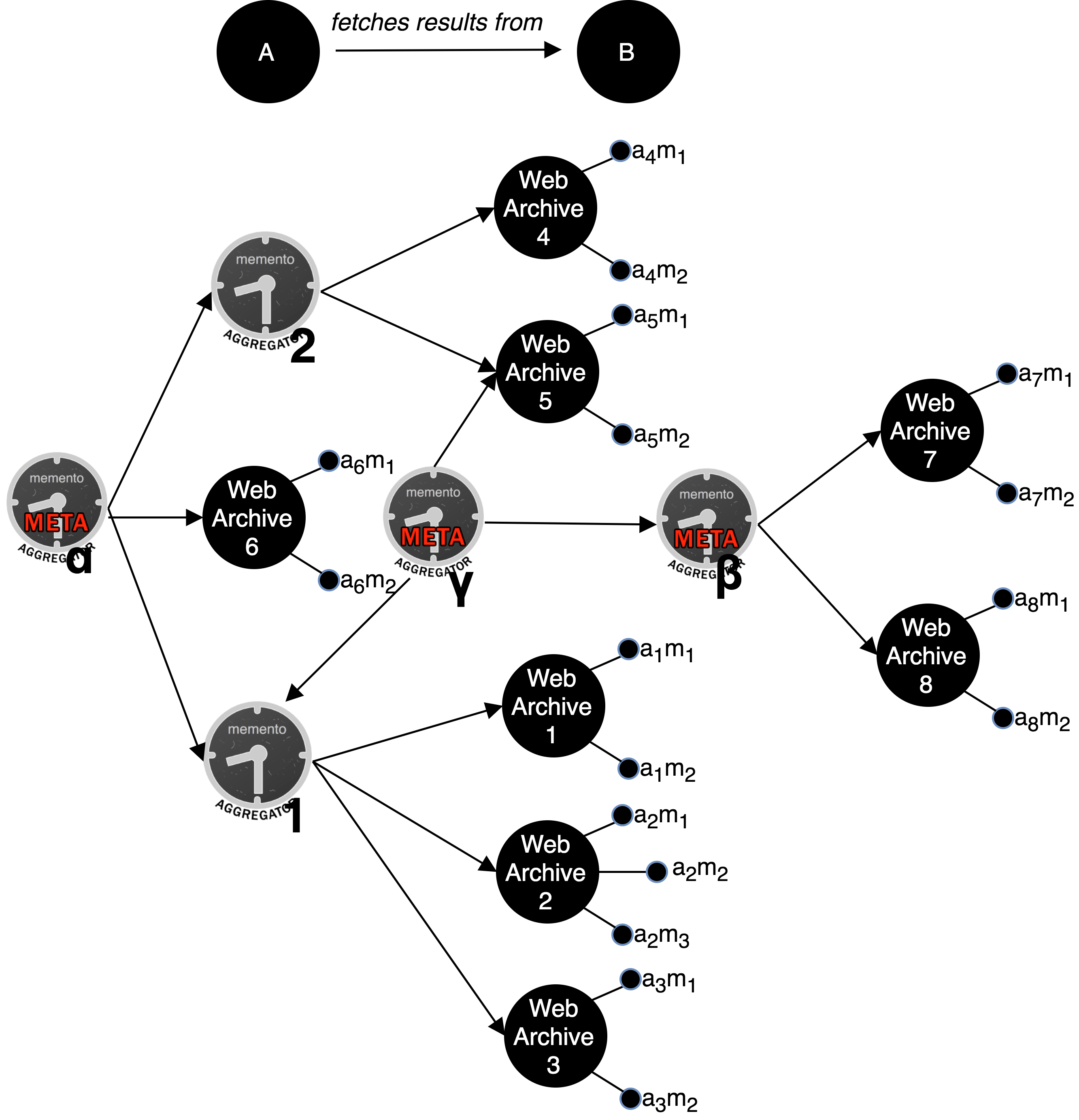} & \specialcell[b]{\begin{tabular}{l l}
\hline
A$_{1\text{...}n}$ & Archive $1$ of $n$ \\
MA$_{1\text{...}n}$ & Memento Aggregator $1$ of $n$ \\
MMA$_{\alpha\text{...}\omega}$ & Memento Meta-Aggregator $1$ of $n$ (denoted using Greek) \\
\blackcircle{3pt}\ a$_x$m$_y$ & Memento of index $y$ from archive of index $x$ \\ \hline
\end{tabular}
\\ \vspace{2.0em} \\
\includegraphics[width=0.4\linewidth]{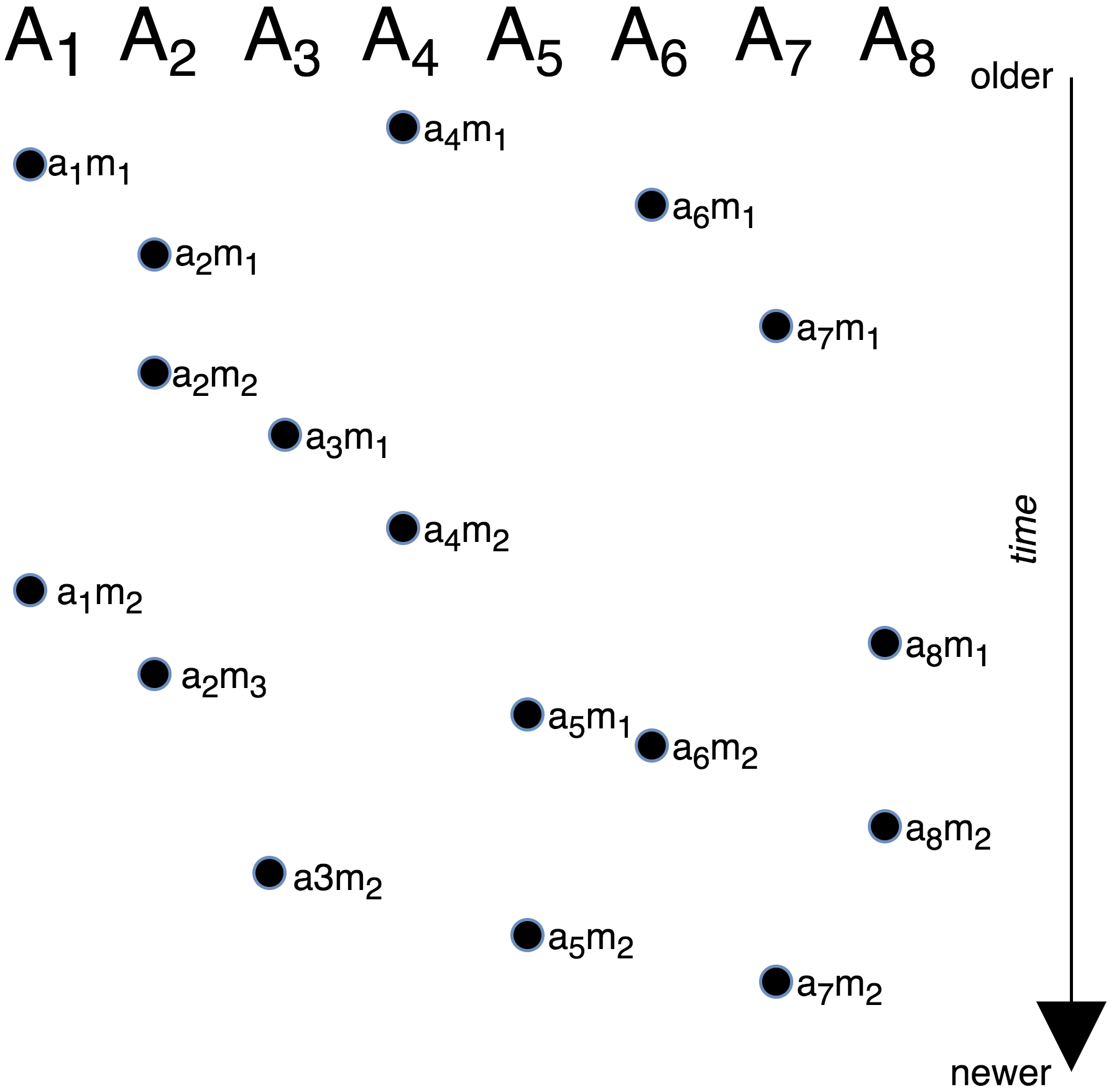} \vspace{4.0em}
}
\end{tabular}\\
\begin{tabularx}{\textwidth}{R p{0.2em} L p{0.2em} L}\hline
\textbf{Mementity} & \arr & \textbf{Abstracted Holdings} & \arr & \textbf{Memento Holdings} \\ \hline
\textit{MA}_1 &  & \{A_1, A_2, A_3\} &  & \{a_1m_1, a_2m_1, a_2m_2, a_3m_1, a_1m_2, a_2m_3, m_3m_2\} \\
\textit{MA}_2 &  & \{A_4, A_5\} &  & \{a_4m1, a_4m2, a_5m_1, a_5m_2\} \\
\textit{MMA}_\alpha &  & \{\textit{MA}_1, \textit{MA}_2, A_6\}\ $\arr$\ \{A_1, A_2, A_3, A_4, A_5, A_6\} & & \{a_4m_1, a_1m_1, a_6m_1, a_2m_1, a_2m_2, a_3m_1, a_4m_2, a_1m_2,a_2m_3, a_5m_1, a_6m_2, a_3m_2, a_5m_2\} \\
\textit{MMA}_\beta & & \{A_7, A_8\} &  & \{a_7m_1, a_8m_1, a_8m_2, a_7m_2\} \\
\textit{MMA}_\gamma & & \makecell[l]{\{\textit{MA}_1, A_5, \textit{MMA}_\beta\}\ $\arr$\ \{A_1, A_2, A_3, A_5, A_7, A_8\}} & & \{a_1m_1, a_2m_1, a_7m_1, a_2m_2, a_3m_1, a_1m_2, a_8m_1, a_2m_3, a_5m_1, a_8m_2, a_3m_2, a_5m_2, a_7m_2\} \\
\end{tabularx}
\end{mdframed}
\caption{Memento Meta-Aggregators may aggregate \URIMs from archives, Memento aggregators, and other MMAs equivalently. Shown is an example of temporally sorted captures as served from an MMA in a variety of permutations in a potentially ad hoc hierarchy.}
\label{fig:mmaHierarchy}
\end{figure*}

A Memento Meta-Aggregator (MMA) serves as a functional superset of a conventional Memento Aggregator (MA). A conventional MA provides access through identifiers to mementos (\URIMsns), TimeGates (\URIGsns), and TimeMaps (\URITsns) from a set of Web archives. An MMA provides the ability to both supplement (Figure~\ref{fig:mmaHierarchy}) and selectively filter the results returned from an MA with \URIMs from additional Web archives at the request of the user or as configured with the MMA (Figure~\ref{fig:mmaScenario}). In a proof-of-concept of this work, we build upon the open source MemGator to introduce these additional roles outside of the scope of the Memento aggregator \entity type to define the MMA \entity type.

Figure~\ref{fig:mmaHierarchy} describes a sample hierarchical relationship of \entities consisting of MMAs, MAs, and Web archive (WAs). When MA${_1}$ receives a request for \URIMs for a \URIRns, for instance, the request is relayed to WA$_1$, WA$_2$, and WA$_3$ for the sets of mementos \{$a_1m_1, a_1m_2$\}, \{$a_2m_1, a_2m_2, a_2m_3$\}, and \{$a_3m_1, a_3m_2$\}, respectively. MA$_1$ is then responsible for combining and temporally sorting the \URIMs then returning the aggregated TimeMap to the requesting user (or \entityns). The temporal ordering within an archive corresponds to the second index ($m$) for convenience in the figure, however, this ordering may not hold between archives. For example, $a_2m_2$ is older than $a_3m_1$ per the temporal ordering diagram on the right side of the figure. The ordering for the mementos contained within the configured archives as requested from various \entities is displayed in the bottom portion of the figure. This figure also shows examples of an MMA obtaining results from multiple MAs (e.g., MMA$_\alpha$ from MA$_1$ and MA$_2$) and even MMAs referring to other MMAs for their results when queried (e.g., MMA$_\gamma$ referring to MA$_1$, WA$_5$, and MMA$_\beta$ with the latter referring to WA$_7$ and WA$_8$). The configuration of MMA$_\beta$ is similar to the relationship of MMA$_{Carol}$ to MMA$_{Alice}$ in Figure~\ref{fig:mmaScenario} where a user may configure an MMA to both refer to a custom set of sources for results as well as reuse the in-place selective filtering of the sources. In this case, MMA$_{Carol}$ would inherit the restriction of MMA$_{Alice}$ of not sending requests for mementos of \url{http://alicesembarassingphotos.net/vacation.html} to Bob's archive.

\begin{figure}
\footnotesize
\begin{mdframed}
\centering
\begin{tabular}{c c c c c c}
\multicolumn{2}{c}{A = Alice's archive} & \multicolumn{2}{c}{B = Bob's archive} & \multicolumn{2}{c}{C = Carol's archive} \\
\multicolumn{3}{c}{I = Internet Archive} & \multicolumn{3}{c}{R = \URIR} \\
\multicolumn{6}{c}{MMA$_X$ = Set of archives sourced for \textit{X}'s MMA for R}\\
\multicolumn{6}{c}{MA = Memento aggregator at \url{mementoweb.org}}
\end{tabular}
\end{mdframed}

\begin{tabular}{l l}
\text{MMA$_{Alice}$=} &
$\begin{cases}
   \{A,B,C\},&\text{``facebook.com''} \in \text{R}\\
   \{A,C\},&\text{``alicesembarassingphotos.net/vacation.html''} \in \text{R}\\
   \{A,B,C,I\},& \text{\textit{otherwise}}
\end{cases}$\\
\text{MMA$_{Bob}$=} &
$\begin{cases}
  \{B,A\}
\end{cases}$\\

\text{MMA$_{Carol}$=}&
$\begin{cases}
   \{C\},&\text{``carolsembarassingphotos.net''} \in \text{R}\\
   \{\text{MMA$_{Alice},MA$}\},& \text{\textit{otherwise}}
\end{cases}$

\end{tabular}
\caption{A Memento Meta-Aggregator is configured to perform selective aggregation.}
\label{fig:mmaScenario}
\end{figure}

Results from other Web archives that are aggregated with the results from an MA may be public non-aggregated Memento-compliant Web archives or private Web archives. We note that a conventional MA is not required to be present to use an MMA, because the aggregation of a static set of public Web archives may be performed by an MMA in a black box banner as if the MMA were identically configured with the same archives as the MA.

An MMA can be configured to return an aggregated TimeMap based on a set of Web archives for which it has been configured or it may be provided a set of archives to query upon request from a client. This abstraction provides a level of extensibility to current Memento aggregators for which the additional functionality may not be appropriate, scalable, or interoperable; however, providing an on-demand set of archives to query is useful in the context of personal Web archiving.

\renewcommand\theadalign{tc}

User-driven specification of aggregation parameters is particularly important for accessing personal Web archives using a Memento aggregator. If a user requests a TimeMap from a conventional Memento aggregator (Figure~\ref{fig:ma}), the aggregator will request the \URIMs from each archive with which the aggregator is configured to communicate. A user may wish to customize, prioritize, or give precedence to the archives queried (as described in Section~\ref{sec:precedence}). If a user hosts an aggregator themselves, the aggregator would need to be reconfigured to prevent requests for certain \URIRs from propagating to certain archives on the basis of \URIRns-archive pairs. Though this may become unwieldy, what follows is a useful example to illustrate where configuring an MMA with a core ruleset prior to considering further user-driven specification would be useful when aggregating personal and public Web archives.

\subsection{MMA Archive Selection}
\label{sec:archivalSelection_mma}
Figure~\ref{fig:mmaScenario} abstracts the following scenario to show how an MMA can perform selective aggregation. Alice archives Web pages she views in her browser using WARCreate \cite{kelly-jcdl12} and replays them using her local Wayback instance within WAIL \cite{berlin-wail}. Bob, Alice's acquaintance, and Carol, Alice's sister, each do the same for their own captures. Alice sets up an MMA (MMA$_{Alice}$) that is configured to request captures from her archive (A), Bob's archive (B), Carol's archive (C), and the Internet Archive (I). For some \URIRsns, like \url{facebook.com}, it may not make sense to aggregate Alice, Bob, and Carol's captures with those from Internet Archive, so she can specify a rule of only aggregating mementos from \{A, B, C\} when those \URIRs are requested\footnote{Note that MMAs do not protect the contents of an archive from being viewed, which is handled by the \entity described in Section~\ref{sec:auth}.}. For other \URIRsns, like \url{alicesembarrasingphotos.net}, Alice may want to prevent exposing the fact that she is looking for certain old captures to Bob and the Internet Archive, but wants to also aggregate captures from Carol's archive, with whom she does not mind exposing the \URIRs requested. She does this by creating another rule to only aggregate from archives \{A,C\} in those cases. By Alice controlling the MMA, she can both pre-configure the set of potential archives queried as well as provide the ability for her, Bob, or Carol to selectively aggregate from the set of archives when requesting captures for a \URIRns. Were Bob uncomfortable with his aggregation requests going to Carol's archive when he used Alice's MMA, he may set up his own MMA (MMA$_{Bob}$) to request captures from only his and Alice's archives without a \URIR filtering scheme like Alice's MMA. Carol also sets up an MMA (MMA$_{Carol}$) that defaults to using Alice's MMA and the \url{mementoweb.org} MA except when requesting \URIRs from \url{carolsembarrassingphotos.net}.

As an endpoint, MMAs may aggregate and request access to captures to private Web archives using a token-based authorization model (e.g., using OAuth \cite{rfc6749} as described further in Section~\ref{sec:auth}). The query may be subsequently routed to an applicable and corresponding Web archive (private or public) after authentication has been established. MMAs may query other MMAs with the expectation that the results returned will be consistent with those from an MA with additional indicators for content beyond the scope of an MA (e.g., a flag for content from a non-aggregated or public archive). In the scenario above, Carol may want additional archives aggregated beyond the default case in Figure~\ref{fig:mmaScenario} so she can utilize the ruleset of Alice's MMA, as well as add filtering rules of her own. The filtering that an MMA performs may not be (and more likely is not) exposed to clients or other MMAs that look to it as a source for \URIMsns. 
Note that in the case of Carol's MMA, there exists a redundancy in that both Alice's MMA and the \url{mementoweb.org} MA will request \URIMs from IA. While Carol's MMA may perform an operation to consolidate duplicates (i.e., a ``UNIQUE'' operation), time may still be wasted waiting for all archived sources to respond to requests to Carol's MMA. Carol may also only want to look to some archives if none, too few, or some other quantifier or qualifier exists in an initial set or series of archives. For advanced querying of this sort, a separate \entity exists and is described in Section~\ref{sec:stargate}.

\subsection{User-driven Archival Specification}

As in the scenario described above, a user may wish to include additional archives in the aggregation process or specify the exclusion of \URIMs from specific archives at the time of the request. The MMA \entity type allows a user to be more descriptive in the results they would like returned compared to a conventional Memento aggregator where only a \URIR and a datetime are specified. Introducing a separate \entity instead of assigning additional roles to the existing Memento aggregator concept provides extensibility while retaining the semantic responsibilities of conventional aggregators and reusing existing infrastructure.

For a user to be able to express additional archives to be aggregated at run time requires both cooperation of the client and recipient to communicate through the same ``protocol''. We accomplish this using a fabricated \texttt{X-More-Archives} HTTP request header, that is consumed by a modified MemGator (serving as an MMA) to supplement the list of archives to be queried (see curl command below). Additional attributes may be specified to the MMA, for instance, if the newly supplied archive requires special handling.

\begin{lstlisting}
curl -H "X-More-Archives: http://myLocalWebArchive/myCollection/timemap/*/"
  "http://mmaHost/timemap/json/http://www.themaneater.com"
\end{lstlisting}

\section{StarGate}
\label{sec:stargate}

Memento TimeGates perform content negotiation in the dimension of datetime (through the Accept-Datetime header) for a \URIR and issue an HTTP 302 redirecting to the appropriate \URIMns. 
This work introduces negotiation with a \entity that serves as an extension to a TimeGate, a StarGate\footnote{``Star'' here refers to common syntax for a wildcard (*)}. A StarGate extends the functionality of a TimeGate with additional content negotiation on other dimensions, such as those described in Section~\ref{sec:additionalTimeMapAttributes}. This broadens archival negotiation beyond the temporal dimension into a range of others. A StarGate also acts as an endpoint to enrich a TimeMap with additional attributes about \URIMsns.

\subsection{Negotiation in Other Dimensions}
\label{sec:negotitationDimensions}
The Prefer HTTP header \cite{rfc7240} provides a basis for content negotiation in other dimensions. Inclusion of the Prefer header requires defining preference in the Vary header of an HTTP response \cite{rfc7240}. Though the specification consists of a registry of preference (of which \texttt{return=minimal} and \texttt{return=representation} are a part), Van de Sompel et al. \cite{preferHeader-wsdlBlog} proposed the extensibility of the definition with \texttt{Prefer} values of \texttt{original-content}, \texttt{original-links}, and \texttt{original-headers}.

\subsection{Authentication and Authorization through Negotiation}
\label{sec:auth}

Figure~\ref{fig:cdxj_new} is an example CDXJ with the \texttt{access} attributes of \texttt{type} and \texttt{token}, which specify for a memento specify a previously established authentication and authorization procedure with a retained token for access persistence. In this initial work, we use an OAuth 2.0 procedure to establish these attributes but the representation is extensible and not coupled to the procedure dynamics.

\begin{figure}
{\small
\begin{enumerate}
\item User requests captures for \URIR from MMA
\item MMA requests \URIR from Public Web Archives $Pu_{1...n}$ and Private Web Archive $Pr_1$\begin{itemize}
  \item $Pu_{1...n}$ each return a respective set of \URIMs $\{\{M_1\}, \{M_2\}, ... \{M_n\}\}$ to MMA
  \item $Pr_{1}$ returns an HTTP 401 and an identifier for an authentication \entity (\URIPns)\end{itemize}
\item MMA returns HTTP 401, \URIPns, and $Pr_1$ identifier to User
\item User sends credentials and \URIR to \URIP
\item \entity at \URIP returns a token to User
\item User requests \URIR again from MMA with token and $Pr_1$ identifier
\item MMA requests \URIR from $Pr_1$ along with token\begin{itemize}
 \item $Pr_1$ returns the set of \URIMs $\{M_{Pr}\}$ to MMA after potentially consulting \entity at \URIP for validity \end{itemize}
\item MMA sorts and transforms $\{\{M_1\}, \{M_2\}, ... \{M_n\},\{M_{Pr}\}\}$ into a TimeMap for \URIR
\item MMA returns TimeMap to User
\end{enumerate}
}
\caption{Abstraction of the authentication to private Web archives follows a flow similar to OAuth 2.}
\label{fig:pwaa_mma_auth}
\end{figure}

Figure~\ref{fig:pwaa_mma_auth} describes the interaction flow of authentication and authorization to a private Web archive. This model uses the model described by OAuth 2.0 wherein the archive from which a capture is being requested takes on the roles of the resource owner and resource server (a fundamental pattern described in the specification), an MMA or user taking on the role of the client, and a \entity at \URIP (an identifier for an authentication \entityns) taking on the role of the authorization server.



\subsection{Endpoint for Archival Aggregate Enrichment}
\label{sec:enrichment}

StarGates are responsible for receiving data about \URIMs and enriching any subsequent TimeMaps. In typical usage of Memento, the TimeGate \entity is not aware of the state of the resources it identifies. Two different approaches can be used to specify additional attributes for \URIMs in a TimeMap: server-driven and client-driven enrichment.

In server-driven enrichment, the StarGate (or some other server-based \entityns) accesses the \URIM for a \URIR in the archive and attempts to acquire the content-based attributes described in Section~\ref{sec:contentBasedAttributes}. These attributes are then retained by the server-based \entity and added inline to TimeMap when the respective original \URIR is requested by a client.

A client-based enrichment approach involves further user interaction of the client accessing a \URIM and a StarGate. A StarGate may provide a \URIM that acts as a proxy to the \URIM that would be requested in conventional Memento usage. By acting as a proxy, a StarGate may setup further communication between the client and StarGate when a \URIM is accessed. This may be accomplished using conventional JavaScript callbacks, a runtime executed service worker (similar to the approach for rerouting at replay-time by Alam et al. \cite{alam-serviceWorker}). 

The client-side approach also allows for distribution of the computation procedure for derived attributes (Section~\ref{sec:derivedAttributes}) with the StarGate acting as an endpoint for the result. As a safeguard to prevent malicious or miscalculated data from being served in a TimeMap, a StarGate may use a consensus model to ensure the accuracy of the result prior to associating it with a \URIMns. As an example, a StarGate may change the \URIM in Figure~\ref{fig:cdxj} from \url{http://web.archive.org/web/19981212013921/http://facebook.com/} to \url{http://stargatehost/calculate/http://web.archive.org/web/19981212013921/http://facebook.com/}. Upon a client accessing the latter URI, the StarGate returns a page to the client with callback information, a key to associate the calculation procedure, and a redirect to the former \URIM with additional embedded JavaScript. For clients that do not support service worker or JavaScript (e.g., curl), accessing the \texttt{stargatehost} URI will provide the same experience the client would receive if accessing the non-proxied \URIMns.

\section{Future Work and Conclusions}
We developed initial prototypes of the MMA\footnote{\url{https://github.com/machawk1/gogator}} (extending MemGator \cite{memgator}) and StarGate\footnote{\url{https://github.com/machawk1/stargate}} as well as integrated a prototype of Mink \cite{kelly2014mink} with the capabilities to interact with other \entities through the dynamics described in this work. Through the implementation of the concepts, \entitiesns, and dynamics described here, users with private Web archives may aggregate their captures both with other private as well as public Web archives to get a better picture of the Web as it was without compromising the information contained in their captures. In the future, we anticipate exploring additional attributes to associate with \URIMs as classified and described in Section~\ref{sec:additionalTimeMapAttributes}. We also anticipate exploring the temporal and spatial ramifications of further supplementing TimeMaps and Link response headers as well as the ramifications on caching and efficient querying and aggregation of the TimeMaps by users and \entities alike. We plan on further exploring archival query precedence models (Section~\ref{sec:precedence}) to consider attributes of archives and their contained mementos in dimensions beyond public-private as well as more complex asynchronous models and short-circuiting techniques.

In this work we laid the foundation for aggregating private and public Web archives. We introduced conceptual Web archiving Memento entities (\entitiesns) to facilitate a hierarchical approach toward aggregation and provided extensible means and methods to consider to further aggregate these captures for a better picture of the Web of the past. We introduced and explored archival precedence and short-circuiting of requests to archival aggregators to allow querying of individual or subsets of archives and to halt if and when a condition is met. We provided the syntax and semantics for enriching Memento TimeMaps with additional attributes to encourage them to be more expressive, particularly as required when aggregating private mementos. We introduced a model to integrate conventional live Web authentication methods with an additional \entity (StarGate) for systematic access control to private captures both from an individual user as well as from other \entitiesns.

\begin{flushleft}
\textbf{Acknowledgements.} This work is supported by the NEH grant HK-50181-14 and IMLS grant RE-33-16-0107-16. We would also like to thank Scott Ainsworth, Sawood Alam, and Shawn Jones for preliminary reviews. Some icons adapted from the work of Agatha Krych and licensed CC-BY-SA 4.0\footnote{\href{https://github.com/machawk1/jcdl2018-artwork}{https://github.com/machawk1/jcdl2018-artwork}}.
\end{flushleft}

\bibliographystyle{ACM-Reference-Format}
\bibliography{hierarchy} 

\end{document}